\documentclass[aps,prd,nofootinbib]{revtex4-2}

\usepackage{amsmath,amssymb}
\usepackage{graphicx}
\usepackage{units}    
\usepackage{wrapfig}
\usepackage[colorlinks=true]{hyperref}     
\usepackage{titlesec}
\usepackage{xspace}
\usepackage{mathtools}
\usepackage{braket}
\usepackage{soul}
\usepackage{comment}
\usepackage{xcolor}
\usepackage{tikz}
\usepackage{pgfplots}
\usepackage{caption}
\usepackage[normalem]{ulem}
\usepackage{float}
\usepackage{array}
\usepackage{appendix}
\usepackage{bm}

\pgfplotsset{compat=1.18}\newcommand{\pathvar}[1]{\ensuremath{\mathcal{D}{#1}}}

\begin{document}
\newcommand{\pathint}[1]{\ensuremath{\int \mathcal{D}{#1}}}
\newcommand{\gpx}{\ensuremath{g_{+x}}}
\newcommand{\gmx}{\ensuremath{g_{-x}}}
\newcommand{\gpy}{\ensuremath{g_{+y}}}
\newcommand{\gmy}{\ensuremath{g_{-y}}}
\newcommand{\kint}{\ensuremath{\int\frac{d^3\vec{k}}{(2\pi)^3}}}
\newcommand{\xint}{\ensuremath{\int d^4 x \ }}
\newcommand{\xyint}{\ensuremath{\int d^4 x d^4 y }}
\newcommand{\brac}[1]{\ensuremath{\langle #1 \rangle}}
\newcommand{\paren}[1]{\ensuremath{\left( {#1} \right)}}
\newcommand{\dphi}{\delta\phi}
\newcommand{\chpx}{\ensuremath{\chi^2_{x}}}
\newcommand{\chmx}{\ensuremath{\chi^2_{x}}}
\newcommand{\chpy}{\ensuremath{\chi^2_{y}}}
\newcommand{\chmy}{\ensuremath{\chi^2_{y}}}
\newcommand{\Hank}[1]{\ensuremath{H_{\nu}^{(1)}(#1)}}
\newcommand{\Hanks}[1]{\ensuremath{H_{\nu}^{*(1)}(#1)}}
\newcommand{\Jank}[1]{\ensuremath{J_{\nu}(#1)}}
\newcommand{\Yank}[1]{\ensuremath{Y_{\nu}(#1)}}
\newcommand{\ekx}[2]{\ensuremath{e^{i\textbf{#1} \cdot \textbf{#2}}}}
\newcommand{\ekxp}[3]{\ensuremath{e^{i\textbf{#1} \cdot (\textbf{#2} - \textbf{#3})}}}
\newcommand{\xv}{\ensuremath{\textbf{x}}}
\newcommand{\kv}{\ensuremath{\textbf{k}}}
\newcommand{\yv}{\ensuremath{\textbf{y}}}
\newcommand{\qv}{\ensuremath{\textbf{q}}}
\newif\ifshowrevisions

\showrevisionsfalse

\ifshowrevisions
  \newcommand{\added}[1]{{\color{blue}#1}}
  \newcommand{\deleted}[1]{{\color{blue}\sout{#1}}}
  \newcommand{\replaced}[2]{{\color{blue}\sout{#1} #2}}
\else
  \newcommand{\added}[1]{#1}
  \newcommand{\deleted}[1]{}
  \newcommand{\replaced}[2]{#2}
\fi

\newcommand{\WC}[1]{#1}
\newcommand{\WCC}[1]{#1}
\newcommand{\addednew}[1]{\textcolor{black}{#1}}
\newcommand{\WCline}[1]{}

\title{Nonlocal Corrections to Scalar Field Effective Action in de Sitter spacetime}
\author{
  Will Cerne$^{1}$, Teruaki Suyama$^{1}$ \\[0.5em]
  $^{1}$Department of Physics, Institute of Science Tokyo, \\
  2-12-1 Ookayama, Meguro-ku, Tokyo 152-8551, Japan \\[0.5em]
  \texttt{cerne.w.57e6@m.isct.ac.jp}
}

\begin{abstract}

We investigate the one-loop effective action for a test scalar field in a general 
Friedmann-Lemaître-Robertson-Walker (FLRW) background, 
specifically focusing on quantum corrections up to the second order in the interaction strength. 
By employing the Schwinger-Keldysh formalism, we derive the equation of motion for the field expectation value, 
which incorporates not only the standard local radiative corrections 
but also novel nonlocal features: a memory term and a stochastic noise term. 
We identify all ultraviolet divergent structures within these nonlocal terms 
and provide a consistent renormalization procedure. 
To analyze the physical impact of these terms, we apply a local approximation under 
the assumption of slowly-varying fields, 
by which the memory term acts as a negative contribution to the drift coefficient.
As a concrete application, we consider a massive $\phi^4$ theory and 
show that these one-loop corrections lead to a suppression of the field variance 
in the infrared regime compared to the tree-level results. 

\end{abstract}

\maketitle
\section{Introduction}

Inflation has become a firmly established paradigm in modern cosmology,
not only because it resolves the shortcomings of the classical Big Bang scenario,
but also because it provides a compelling mechanism for generating the primordial
density fluctuations observed today \cite{Baumann2009,Planck2018Inflation}.
High-precision measurements of the Cosmic Microwave Background (CMB),
together with large-scale structure (LSS) surveys, have confirmed key predictions
of inflation with remarkable accuracy \cite{ACTDR6Extended,DESI2024VI}. Furthermore, a variety of forthcoming observational programs ranging from next-generation CMB missions to 21-cm surveys and gravitational-wave detectors promise to probe the primordial fluctuations with even greater precision \cite{abitbol2019simons,litebird2023probing,kawamura2021current}.

These advances are expected to significantly improve our ability to constrain 
the inflaton potential and, more generally, 
the dynamics of the fields that drove inflation. 
To fully exploit this observational progress, it is essential to have a theoretically robust understanding of the inflaton's dynamics. 
In standard treatments, the background dynamics of the inflaton are obtained from the classical equations of motion derived from the tree-level action \cite{Boyanovsky_2006}. 

However, in principle these dynamics should be determined by the effective action, which incorporates quantum effects arising from self-interactions of the field 
as well as its coupling to gravity \cite{weinbergQFT2}.  A proper evaluation of these quantum corrections is therefore an indispensable step toward establishing a more precise connection between inflationary theory and observations.
As a first step toward this goal, the present paper focuses on a simpler 
but technically important system: a test scalar field evolving on a fixed inflationary background. 

Studying a test field enables a clean analysis of quantum effects because 
only the scalar-field fluctuations need to be quantized, 
whereas treating the inflaton itself would require a simultaneous quantization 
of metric perturbations, which makes the calculation significantly more involved. 
Despite being a simplified setup, the test-field dynamics already exhibit nontrivial quantum corrections that provide key insights into how such effects 
may influence the inflaton case. 

Quantum corrections to the dynamics of scalar fields in an inflationary background have been studied in several earlier works \cite{Glavan_2023, Vicentini_2019, Markkanen_2018, Herranen_2014, Janssen_2009, Markkanen_2012}. In particular, Boyanovsky et al. \cite{Boyanovsky_2006} analyzed the equation of motion by applying the one-loop (and first-order in interaction) approximation directly to the field equation. 
Their results revealed corrections to the effective potential but were restricted to leading order in the interaction strength. Herranen  \cite{Herranen_2015} then extended their work by including metric fluctuations in the system. Quantum effects appear mostly through the quantization of the scalar field fluctuations from the variance $\langle \delta \phi^2 \rangle$.

In contrast, the approach taken in this work is based on computing the effective action using the in-in (Schwinger--Keldysh) formalism \cite{schwinger1961brownian}, 
which is the appropriate framework for obtaining expectation values in time-dependent backgrounds. 

Markkanen et al. \cite{Markkanen_2012} computes the fully renormalized one loop effective action for a scalar field in the FLRW spacetime. However, the field is held constant which results in a standard-form local effective potential.

In this paper, we evaluate the one-loop effective action for a minimally 
coupled test scalar field in an FLRW background, 
expanding the action perturbatively up to second order in the interaction.
As we show, new physical effects emerge at this second order: in addition to corrections to the effective potential, 
we find a memory term and an additional colored noise term originating from the imaginary part of the effective action. 
These features are absent at first order and have not been derived previously 
from the effective-action perspective in an inflationary background.
Specializing to the de Sitter limit, we renormalize the effective action and derive the equation of motion from this action.
This equation exhibits qualitatively new structures compared to previous results: the drift and noise terms both receive nonlocal, interaction-induced corrections that cannot be captured solely by an effective potential. 

As a case study comparison with regular stochastic inflation, we modify the equation of motion by removing the quantum corrections from the IR regime by hand and inserting standard white Gaussian noise with amplitude $\frac{H}{2\pi}$. We analyze these corrections for a massive quartic
potential and quantify their effects on stochastic expectation values and on the equilibrium probability distribution.
The results presented here therefore clarify, within a controlled approximation, how non local effects alter the stochastic dynamics of scalar fields in an expanding universe. 

Since the inflaton constitutes a special but more complicated case of such fields, the present analysis provides an essential step toward a more complete effective-action treatment of inflationary dynamics.\\
Throughout this paper, we employ natural units, $\hbar = c = 1$, and use the metric convention $\eta_{\mu\nu} = \mathrm{diag}(1,-1,-1,-1)$.

\section{One-Loop Corrections in the Effective Action of the Test Field: The General Case}

The action of a real scalar field in the FLRW background in a potential $V(\phi)$ can be expressed as

\begin{equation}
S[\phi] = \xint a^3(t) \; \left[ \frac{1}{2}g^{\mu\nu}\partial_{\nu}\phi \partial_{\mu}\phi - \frac{1}{2}m^2\phi^2 - V(\phi)\right].
\end{equation}

The mass term has been separated from the potential in order for $V(\phi)$ to represent only the interactions of the field. From here, the field is separated into a mean field and fluctuations about the mean field.

\begin{equation}
\phi = \Phi + \chi  \; ; \, \, \Phi  = \langle \phi \rangle \; , \; \langle \chi \rangle= 0.
\end{equation}

Then, Taylor expanding about the fluctuations results in:

\begin{equation}
S = \xint a^3(t) \left[\frac{1}{2}\partial^{\mu}\left(\Phi + \chi\right) \partial_{\mu}\left(\Phi + \chi\right)  - \frac{1}{2}m^2\phi^2 - V(\Phi) - V'(\Phi)\chi - \frac{1}{2}V''(\Phi)\chi^2 - \frac{1}{6}V'''(\Phi)\chi^3 + \dots \right]
\end{equation}

The effective action is a quantity that can be used to calculate the equation of motion that the quantum expectation value of the field obeys \cite{weinbergQFT2, coleman1973radiative}. 
In the present case, since we study the quantum expectation value of the mean field including radiative corrections, to properly define the effective action one must use the in-in formalism. This is also known as the Schwinger-Keldysh formalism \cite{schwinger1961brownian, keldysh2024diagram}. This results in two path integration variables, denoted $\chi_+$ and $\chi_-$. The plus indicates that the integration is forward in time and the minus indicates that the integration is backward in time.

The effective action $\Gamma$ can be computed as a summation over one particle irreducible (1PI) diagrams as follows:

\begin{equation}
\exp \left( i\Gamma \left[ \Phi_+, \Phi_- \right]\right) = \int_{\text{1PI}} \pathvar{\chi_+} \pathvar{\chi_-} e^{iS\left[\Phi_+ + \chi_+ \right] - iS\left[\Phi_- + \chi_- \right]}
\end{equation}

This path integral is not analytically analyzable. 
In this paper, we evaluate the effective action at the one-loop level, which can be done by
Taylor expanding the integrand about the fluctuations $\chi_\pm$. 
To do this, we note that the first-order terms in $\chi$ do not contribute to 1PI diagrams, 
and the third- and higher-order terms in $\chi$ do not contribute to the effective action at one-loop order.
Then, the one-loop effective action can be rewritten as
\begin{equation}
\label{1-loop-action}
\begin{aligned}
\exp{(i\Gamma_{\text{1-loop}}}\left[ \Phi_+, \Phi_- \right]) = &\int_{\text{1PI}} \pathvar{\chi_+} \pathvar{\chi_-} e^{iS_0\left[\chi_+ \right] - iS_0\left[\chi_- \right]-\frac{i}{2} \xint a^3(t) g(\Phi_+(x))\chi_+^2(x)  + \frac{i}{2} \xint a^3(t) g(\Phi_-(x))\chi_-^2(x)}
\end{aligned}
\end{equation}
where $S_0$ is the tree-level action without the interaction, given by
\begin{equation}
S_0[\phi] \equiv \int d^4x \; a^3(t) \biggr[ \frac{1}{2}\partial^{\mu}\phi \partial_{\mu}\phi - \frac{1}{2}m^2\phi^2 \biggr],
\end{equation}
and $g(\Phi(x)) \equiv  V''(\Phi(x))$ is an effective coupling constant.

\added{The quadratic term $g(\Phi(x))\chi^2/2$ can be organized in two equivalent ways. It is a spacetime-dependent contribution to the fluctuation mass, and it may therefore be included directly in a dressed propagator. Alternatively, after choosing $S_0$ as the free action, the same term may be treated perturbatively as a two-point mass insertion. These descriptions correspond to a resummed and a fixed-order organization of the same quadratic Green function; they are not different physical assumptions.}

\WCC{In principle, the path integral (\ref{1-loop-action}) is Gaussian with an effective mass given by $m^2 + g(\Phi(x))$. In this case, one could define the propagator of the field theory $\boldsymbol{\Delta}$ as
\begin{equation}
\left[
\begin{pmatrix}
-\Box_x-m^2 & 0 \\
0 & \Box_x+m^2
\end{pmatrix}
-
\begin{pmatrix}
g(\Phi_+(x)) & 0 \\
0 & -g(\Phi_-(x))
\end{pmatrix}
\right]
\boldsymbol{\Delta}(x,x')
=
\frac{\delta^D(x-x')}{\sqrt{-g(x)}}
\times \textbf{1},
\label{eq:dressed-propagator-matrix}
\end{equation}
where the propagator is to be understood in the in-in formalism as a 2x2 matrix of propagators. In such cases, the effective action at the 1-loop order can be written schematically as
\begin{equation}
    \Gamma_{\text{1-loop}} = \frac{i}{2}\text{Tr} \ln \left(\frac{1}{\textbf{G}_0^{-1}-\textbf{g}}\right)
\label{eq:1loopexact}
\end{equation}
where $\boldsymbol{\Delta} = \left(\textbf{G}_0^{-1}-\textbf{g}\right)^{-1}$ is defined from Eq.~(\ref{eq:dressed-propagator-matrix}) as the inverse of the free propagator $\textbf{G}_0$ minus the interaction $V''\rightarrow g$. For constant fields $\Phi$, this expression can be evaluated analytically and gives a local contribution to the potential, which would be the curved space analog of the Coleman-Weinberg 1-loop potential.
However, in the present case the field $\Phi(x)$ depends non-trivially on the spacetime point $x$, and thus the dressed propagator in Eq.~(\ref{eq:dressed-propagator-matrix}) acquires this dependence. In this case, Eq.~(\ref{eq:1loopexact}) cannot be evaluated exactly, and must be expanded in powers of the interaction $\textbf{g}$. 
}

\added{Equivalently, the same dressed propagator has the formal Dyson expansion
\begin{equation}
\boldsymbol{\Delta}
=
\textbf{G}_0
+
\textbf{G}_0\,\textbf{g}\,\textbf{G}_0
+
\textbf{G}_0\,\textbf{g}\,\textbf{G}_0\,\textbf{g}\,\textbf{G}_0
+\cdots .
\label{eq:mass-insertion-dyson}
\end{equation}
Thus, $g=V''(\Phi)$ is a spacetime-dependent contribution to the quadratic mass operator, while treating it as a two-point mass insertion specifies a perturbative organization of that operator. If all insertions are retained, the Dyson series resums to the dressed propagator; at finite order it is an approximation whose validity requires the omitted higher orders to remain small. The perturbative treatment of a background-dependent mass contribution is used explicitly in peer-reviewed nonequilibrium field-theory calculations. Boyanovsky \emph{et al.} treat a time-dependent quadratic background term perturbatively in an amplitude expansion, while also discussing its absorption into a time-dependent fluctuation mass \cite{Boyanovsky1995Dissipation}. In the electroweak VEV-insertion approximation, the spacetime-dependent part of the mass is likewise treated as a perturbation \cite{Postma2021VIA,PostmaVanDeVis2020VIA,RiusSanz2000HiggsInsertion}. Ref.~\cite{Postma2021VIA} also compares equivalent bookkeeping choices within that mass-insertion expansion; we use it here as a precedent for the perturbative treatment itself, not as a claim that a finite truncation is identical to the unexpanded dressed propagator. Resummed treatments can be preferable when higher insertions become important \cite{PostmaVanDeVisWhite2022VIA}. In the present work, we restrict the expansion in $g$ to the weak-interaction regime and retain terms through second order.}

\WCC{In this case,} up to second order in $g$ Eq.~(\ref{1-loop-action}) becomes 
\begin{equation}
\begin{aligned}
i \Gamma_{1-\text{loop}}\left[ \Phi_+, \Phi_- \right] =& \int_{\rm 1PI-\text{connected}} \pathvar{\chi_+} \pathvar{\chi_-} e^{iS_0\left[\chi_+ \right] - iS_0\left[\chi_- \right]} \\
& \; \times \biggr(1-\frac{i}{2} \xint a^3(t) \gpx\chi_+^2(x)  + \frac{i}{2} \xint a^3(t) \gmx\chi_-^2(x) \\
& - \frac{1}{8} \xyint  \; a^3(t_x)a^3(t_y)\left( \gpx\chi_+^2(x) - \gmx\chi_-^2(x)\right)\left(\gpy\chi_+^2(y) - \gmy\chi_-^2(y)\right) \biggr),
\end{aligned}
\end{equation}
where $ g_{\pm x}\equiv g(\Phi_\pm(x)) $.

From the above expression, the part that is first order in $g$ is given by:
\begin{equation}
\label{1-loop-1st-g}
\begin{aligned}
\Gamma_{1-\text{loop}}^{(1)}\left[ \Phi_+, \Phi_- \right]= -\frac{1}{2}\pathint{\chi_+} \pathvar{\chi_-} e^{iS_0\left[\chi_+ \right] - iS_0\left[\chi_- \right]} \left( \xint a^3(t) \left( \gpx\chi_+^2(x) - \gmx\chi_-^2(x) \right) \right).
\end{aligned}
\end{equation}
We define $\brac{\cdots}$ by
\begin{equation}
    \brac{\cdots} = \int \pathvar{\chi_+}\pathvar{\chi_-} \; \left( \cdots \right) e^{iS_0\left[\chi_+ \right] - iS_0\left[\chi_- \right]}.
\end{equation}
Then, noting that $\brac{\chi_+^2(x)} = \brac{\chi_-^2(x)} \equiv \brac{\chi^2(x)}$, 
Eq.~(\ref{1-loop-1st-g}) simplifies to
\begin{equation}
\Gamma_{1-\text{loop}}^{(1)}\left[ \Phi_+, \Phi_- \right]=  -\frac{1}{2}\xint a^3(t)\brac{\chi^2(x)}\left( \gpx - \gmx \right).
\end{equation}
At this order, $\brac{\chi^2(x)}$ is the propagator (at the same spacetime point) of a massive free field in the FLRW spacetime and the state is chosen to be the Bunch-Davies vacuum state.

The terms second order in $g$ are 
\begin{equation}
\label{1-loop-2nd-g}
\begin{aligned}
\Gamma_{1-\text{loop}}^{(2)} \left[ \Phi_+, \Phi_- \right]= &\frac{i}{8}\pathint{\chi_+} \pathvar{\chi_-} e^{iS_0\left[\chi_+ \right] - iS_0\left[\chi_- \right]} \xyint  \; a^3(t_x)a^3(t_y)\left( \gpx\chi_+^2(x) - \gmx\chi_-^2(x)\right)\left(\gpy\chi_+^2(y) - \gmy\chi_-^2(y)\right) \\
 = &\frac{i}{8} \xyint \; a^3(t_x)a^3(t_y)\left(\gpx\gpy\brac{T\chi^2(x)\chi^2(y)} + \gmx\gmy\brac{\bar{T}\chi^2(x)\chi^2(y)} -2\gmx\gpy\brac{\chi^2(x)\chi^2(y)} \right),
\end{aligned}
\end{equation}
where $T$ and ${\bar T}$ are the time-ordering and anti-time-ordering operators respectively.
Since these expectation values are for free fields, Wick's Theorem can be used to break up the four-point correlators into propagators as
\begin{equation}
\brac{{\cal O} \chi^2(x)\chi^2(y)}=2 {\brac{{\cal O} \chi(x)\chi(y)}}^2+
{\brac{\chi^2(x)}} {\brac{\chi^2(y)}},
\end{equation}
where the operator ${\cal O}$ is $T$, ${\bar T}$, or the identity operator.
With this decomposition and using the fact that the second term ${\brac{\chi^2(x)}} {\brac{\chi^2(y)}}$ in the decomposition corresponds to disconnected diagrams, 
Eq.~(\ref{1-loop-2nd-g}) becomes
\begin{equation}
\label{1-loop-2nd-g-2}
\begin{aligned}
\Gamma_{1-\text{loop}}^{(2)} \left[ \Phi_+, \Phi_- \right]=\frac{i}{4} \xyint \; a^3(t_x)a^3(t_y)\biggr( \gpx\gpy\brac{\text{T}\chi_{x}\chi_{y}}^2 +\gmx\gmy \brac{\bar{\text{T}}\chi_{x}\chi_{y}}^2 -2\gmx\gpy \brac{\chi_{x}\chi_{y}}^2 \biggr),
\end{aligned}
\end{equation}
where $\chi_x \equiv \chi(x)$. 

\replaced{As is well known, and as is also the case here,
$\Gamma_{1-\text{loop}}^{(2)}$ contains an imaginary part,
indicating that the system under consideration is dissipative.}{At second order, the CTP effective action separates into a real retarded response term and an imaginary term quadratic in the CTP difference field. The former gives the memory (dissipative-response) contribution to the physical mean-field equation, while the latter is associated with noise.} 
To see this explicitly, we first perform a change of field variables as
\begin{equation}
\Phi_{\pm}(x) = \Phi(x) \pm \frac{1}{2}\Phi_{\Delta}(x).
\end{equation}
These new variables are motivated by technical reasons. 
The equation of motion of the expectation value of the field is obtained 
by imposing $\Phi_+=\Phi_-$ after
the Euler-Lagrange equations from the effective action are derived.
In terms of the new variables, this manipulation amounts to setting $\Phi_\Delta =0$ in
the functional derivative of the effective action with respect to $\Phi$.
Also, the factor of $\frac{1}{2}$ is included such that the tree-level contribution to the action is $\int d^4x~\frac{\delta S[\Phi]}{\delta \Phi} \Phi_{\Delta}$ with no extra coefficients.
In terms of the new variables, the functions $g_\pm$ become
\begin{equation}
g_{\pm x} = g(\Phi(x)\pm \frac{1}{2}\Phi_{\Delta}(x)) \approx g(\Phi(x)) \pm \frac{1}{2}g'(\Phi(x))\Phi_{\Delta}(x) + \frac{1}{8}g''(\Phi(x))\Phi_{\Delta}^2(x).
\end{equation}

With this, Eq.~(\ref{1-loop-2nd-g-2}) becomes
\begin{equation}
\label{1-loop-2nd-g-3}
\begin{aligned}
\Gamma_{1-\text{loop}}^{(2)} \left[ \Phi, \Phi_\Delta \right] =&- \int d^4x d^4y \; a^3(t_x) a^3(t_y) g'(\Phi(x)) g(\Phi(y)) \theta(t_x-t_y)\Phi_{\Delta}(x) \times \text{Im}\biggr( \brac{\chi(x)\chi(y)}^2 \biggr) \\
& + \frac{i}{4}\int d^4x d^4y \; a^3(t_x) a^3(t_y) g'(\Phi(x)) g'(\Phi(y)) \Phi_{\Delta}(x)\Phi_{\Delta}(y) \times \text{Re}\biggr( \brac{\chi(x)\chi(y)}^2 \biggr), 
\end{aligned}
\end{equation}
where $\theta(x)$ is the step function.

\added{The first line of Eq.~(\ref{1-loop-2nd-g-3}) is real and explicitly retarded because of $\theta(t_x-t_y)$. We therefore refer to it as the \emph{retarded memory (dissipative-response) term}. The second line is imaginary and quadratic in $\Phi_\Delta$; it is the term associated with fluctuations/noise in the CTP effective action. This distinction is useful because a real retarded response does not require an imaginary coincident propagator.}

\replaced{The second line in Eq.~(\ref{1-loop-2nd-g-3}), which is non-zero in general and contributes as an imaginary part to the effective action,
is known to be interpreted as noise (see \cite{morikawa1986classical}).}{The imaginary second line of Eq.~(\ref{1-loop-2nd-g-3}) is interpreted as noise (see \cite{morikawa1986classical}).}
In order to demonstrate this is actually the case in our analysis,
we introduce a function $N(x,y)$ by
\begin{equation}
\label{noise-correlation}
N(x,y) \equiv \frac{1}{4}a^3(t_x) a^3(t_y) V'''(\Phi(x))V'''(\Phi(y))\text{Re}\biggr( \brac{\chi(x)\chi(y)}^2\biggr),
\end{equation}
by which we have
\begin{equation}
{\rm Im}\Gamma_{1-\text{loop}}^{(2)} \left[ \Phi, \Phi_\Delta \right] =\int d^4x d^4y ~
\Phi_\Delta (x) N(x,y) \Phi_\Delta (y).
\end{equation}
This function is positive-definite in the sense that 
\begin{equation}
(\phi, N\phi) \equiv  \int d^4x d^4y \; \phi(x) N(x,y) \phi(y) \geq 0
\end{equation}
for any functions $\phi$.
This can be verified by introducing a Hermitian operator $A$ by
\begin{equation}
A=\frac{1}{2\sqrt{2}} \int d^4x~a^3 (t_x) V'''(\Phi(x)) \phi(x) \chi^2 (x)
\end{equation}
and writing $(\phi, N\phi)$ as
\begin{equation}
(\phi, N\phi)=\langle A^2 \rangle-\langle A \rangle^2=\langle {(A-\langle A \rangle )}^2
\rangle,
\end{equation}
which is positive.
When this positivity holds, there is an identity given by
\begin{equation}
\exp \left( - (\Phi_\Delta,N\Phi_\Delta) \right) = \int \pathvar{\xi} \exp \left( -\frac{1}{4} \int d^4xd^4y \; \xi(x) N^{-1}(x,y) \xi(y) + i \int d^4x \; \xi(x)\Phi_\Delta(x) \right).
\end{equation}
Using this identity, the effective action becomes
\begin{equation}
\exp \left( i\Gamma_{\text{1-loop}} \right) = \int \pathvar{\xi} \; \mathcal{P}[\xi] 
\exp \left( i{\rm Re} \Gamma_{\text{1-loop}} 
[\Phi,\Phi_\Delta] + i\int d^4x \; \xi(x)\Phi_\Delta(x)\right),
\end{equation}
where we have defined the functional $\mathcal{P}[\xi]$ by
\begin{equation}
\mathcal{P}[\xi] \equiv \exp \left( -\frac{1}{4} \int d^4xd^4y \; \xi(x) N^{-1}(x,y) \xi(y) \right).
\end{equation}
The positive nature and Gaussian form of $\mathcal{P}[\xi] $ suggest that 
$\xi$ is a Gaussian random variable (noise) interacting with $\Phi_\Delta$ by the term
$\xi \Phi_\Delta$ \cite{morikawa1986classical}.
The two-point function of $\xi$ is given by
\begin{equation}
\brac{\xi(x)\xi(y)} = 2N(x,y).
\end{equation}

Having rewritten the imaginary part of the effective action as the noise term,
the effective action under the current approximation now reads
\begin{equation}
\begin{aligned}
\label{effective-action-result}
\Gamma_{\text{1-loop}} \left[\Phi(x),  \Phi_{\Delta}(x)\right]& =\;  \int d^4x \; \frac{\delta S[\Phi]}{\delta \Phi}\Phi_{\Delta}(x) \\ 
& - \frac{1}{2} \xint a^3(t)V'''(\Phi(x))\Phi_{\Delta}(x)\brac{\chi^2(x)}\\
& -\int d^4x d^4y \; a^3(t_x) a^3(t_y) V'''(\Phi(x)) V''(\Phi(y)) \theta(t_x-t_y)\Phi_{\Delta}(x) \times \text{Im}\biggr( \brac{\chi(x)\chi(y)}^2 \biggr) \\
& + \int d^4x \; \Phi_{\Delta}(x) \; \xi(x) + \mathcal{O}(\Phi_{\Delta}^3),
\end{aligned}
\end{equation}
where $g$ has been replaced with $V''(\Phi)$.
The equation of motion of $\Phi$ can be obtained as the Euler-Lagrange equation
for $\Phi_\Delta$;
\begin{equation}
\frac{\delta \Gamma_{\text{1-loop}}}{\delta \Phi_\Delta} \biggr|_{\Phi_\Delta=0} = 0.
\end{equation}
Using the explicit form of $\Gamma_{\text{1-loop}}$, the equation of motion 
of $\Phi$ becomes
\begin{equation}
\label{EOM-Phi}
(\Box +m^2)\Phi(x) + V'(\Phi (x)) + \frac{1}{2}V'''(\Phi(x))\brac{\chi^2(x)} + V'''(\Phi(x))\int_{-\infty}^{t_x}\int d^3y \; a^3(t_y) V''(\Phi(y)) \text{Im}\biggr( \brac{\chi(x)\chi(y)}^2 \biggr) = \xi(x).
\end{equation}

\added{It is worth stressing that the retarded memory term in Eq.~(\ref{EOM-Phi}) is fully compatible with a real coincident dressed propagator. If $V''(\Phi)$ is absorbed into the exact quadratic operator, $i\Delta^{++}(x,x)$ is real in the physical CTP limit, but it is generally a nonlocal functional of the earlier background because the propagator solves an equation with the spacetime-dependent coefficient $V''(\Phi(x'))$. Expanding that same dressed propagator according to Eq.~(\ref{eq:mass-insertion-dyson}) exposes this past dependence as the real retarded integral in Eq.~(\ref{EOM-Phi}). Thus, the perturbative organization does not create an imaginary part of the coincident propagator; it makes the causal memory contained in the dressed Green function explicit.}

Closely related structures for non-Markovian memory and colored-noise structures arise when tracing out spectator fields using the Schwinger-Keldysh formalism, see \cite{Wu_2007, Lee_2011}. In contrast, the present equation is derived directly from the one-loop effective action for the mean field.
The first two terms on the left-hand side are classical ones.
The third term, which is UV divergent, gives a radiative correction to the 
tree-level potential.
As we will show in the next section, it contains a finite and time dependent
component which depends on the behavior of $a(t)$ and cannot be renormalized into the tree-level potential.
Thus, this finite component gives a physical correction to the classical equation
of motion.
Contrary to the other terms, the fourth term, which is also UV divergent, 
depends not only on the field 
value in the vicinity of the spacetime point under consideration 
but also on the field values at distant locations. 
Thus, this term possesses a non-local character.
Because the upper limit of the time integration is $t_x$, only past information
affects the dynamics of $\Phi$ (memory term).
Furthermore, from the identity, $\text{Im}\left(\brac{\chi(x)\chi(y)}^2\right) =\frac{1}{2i} \langle \{ \chi (x), \chi (y) \} \rangle \langle [ \chi (x), \chi (y) ] \rangle$,
the integrand of the memory term vanishes when $x$ and $y$ are spacelike separated because the commutator of two operator fields at spacelike separated points is zero. 
Thus, the memory term contains only information in the past light cone
and respects causality.

It is important to note that the derivation of the equation of motion (\ref{EOM-Phi}) has been purely general in the sense that it is insensitive to the specific potential. Any general potential that allows perturbative expansion satisfies, to one-loop order, the equation of motion given above. 
Also, the derivation of Eq.~(\ref{EOM-Phi}) is true for any FLRW metric, since none of the specific properties of $a(t)$ were used in its derivation. 
However, the renormalization of the above terms will depend on the behavior of $a(t)$.

\section{Renormalization}

\subsection{Renormalization of the First Order Term}
\label{Ren-1st}
The term of our current interest is the first-order term given by
\begin{equation}
\Gamma_{\text{1-loop}}^{(1)}[\Phi, \Phi_\Delta] \equiv -\frac{1}{2}\int d^4 x \, a^3(t) V'''(\Phi(x)) \langle \chi^2(x) \rangle \Phi_\Delta(x).
\end{equation}

The purpose of this subsection is to clarify the properties of the UV divergence arising 
from $\langle \chi^2(x) \rangle$. 
At the level of our approximation, this can be evaluated by treating $\chi$ 
as a free scalar field on the FLRW spacetime, for which
the operator $\chi$ in the Heisenberg picture can be expressed as
\begin{equation}
\chi(x) = \int \frac{d^3 p}{(2\pi)^3 2} \left( a(\vec{p}) e^{i \vec{p}\cdot \vec{x}} u_p(t) + a^\dagger(\vec{p}) e^{-i \vec{p} \cdot \vec{x}} u_p^*(t) \right),
\end{equation}
where $a(\vec{p})$ and $a^\dagger(\vec{p})$ are annihilation and creation operators satisfying
\begin{equation}
[a(\vec{p}), a^\dagger(\vec{q})] = \delta(\vec{p}-\vec{q}),
\end{equation}
and $u_p(t)$ is the mode function which is a solution of the differential equation
\begin{equation}
\ddot{u}_p + 3 H \dot{u}_p + \left(\frac{p^2}{a^2} + m^2 \right) u_p = 0,
\end{equation}
with the normalization condition
\begin{equation}
u_p \dot{u}_p^* - u_p^* \dot{u}_p = \frac{i}{a^3(t)},
\end{equation}
and the positive frequency condition
\begin{equation}
u_p \sim \exp \left( -i \int^t \omega_p(s) \, ds \right), \quad \omega_p \equiv \sqrt{\frac{p^2}{a^2} + m^2}.
\end{equation}

With this expression, $\langle \chi^2 \rangle$ for the vacuum state $|0\rangle$ (i.e., $a(\vec{p}) |0\rangle = 0$) becomes
\begin{equation}
\langle \chi^2(x) \rangle = \int \frac{d^3 p}{(2\pi)^3} |u_p|^2.
\end{equation}
In the UV regime, the mode oscillates much faster than the background varies. 
For such a case, we can derive the approximate solution using the WKB method.
It is given by
\begin{equation}
|u_p|^2 = \frac{1}{2 a^2 p} \left[ 1 - \frac{a^2 m^2}{2 p^2} + \frac{a^2}{2 p^2} \left( H^2 + \frac{\ddot{a}}{a} \right) \right] + O\left( \frac{1}{p^5} \right).
\end{equation}
Introducing a physical cutoff $\Lambda$, the UV divergent part of $\langle \chi^2 \rangle$ 
then becomes
\begin{equation}
\langle \chi^2(x) \rangle_{\text{UV}} = \frac{\Lambda^2}{8 \pi^2} - \frac{1}{8 \pi^2} \left( m^2 - \frac{R}{6} \right) \ln \frac{\Lambda}{\WC{\mu}},
\end{equation}
where $R = 6 (H^2 + \ddot{a}/a)$ is the Ricci scalar for the FLRW metric.
Thus, we decompose
\begin{equation}
\langle \chi^2(x) \rangle = \langle \chi^2(x) \rangle_{\text{Min}} + \left( \langle \chi^2(x) \rangle - \langle \chi^2(x) \rangle_{\text{Min}} \right),
\end{equation}
\WCC{with $\brac{\chi^2(x)}_{\text{Min}} \equiv \frac{\Lambda^2}{8\pi^2} - \frac{m^2}{8\pi^2}\ln\frac{\Lambda}{\mu}$ being the free massive propagator at coincidence in Minkowski space.}

Using this,

\begin{equation}
\langle \chi^2(x) \rangle - \langle \chi^2(x) \rangle_{\text{Min}} = \frac{R}{48 \pi^2} \ln \frac{\Lambda}{\WC{\mu}} + A(x),
\end{equation}
where $A(x)$ is UV-finite. Then, the first-order term becomes
\begin{align}
\label{1-loop-1st-order}
\Gamma_{\text{1-loop}}^{(1)}[\Phi, \Phi_\Delta] &= -\frac{1}{2}  \int d^4 x \, a^3(t) V'''(\Phi(x)) \langle \chi^2(x) \rangle_{\text{Min}} \Phi_\Delta(x) -\frac{1}{2}  \int d^4 x \, a^3(t) V'''(\Phi(x)) \left( \frac{R}{48 \pi^2} \ln \frac{\Lambda}{\WC{\mu}} \right) \Phi_\Delta(x) \notag \\
&\quad -\frac{1}{2}  \int d^4 x \, a^3(t) V'''(\Phi(x)) A(x) \Phi_\Delta(x).
\end{align}
The first term on the right-hand side contributes to the renormalized potential
\begin{equation}
V_R(\Phi) \equiv V(\Phi) + \frac{1}{2} V''(\Phi) \langle \chi^2(x) \rangle_{\text{Min}}.
\end{equation}
The second term, which is proportional to the Ricci scalar $R$, represents the non-minimal 
coupling between the scalar field and $R$.
Although we started from the assumption that the scalar field is minimally coupled to gravity,
the non-minimal coupling term automatically arises from the radiative correction.
Alternatively, if the tree-level Lagrangian contains a non-minimal coupling term 
of the form $\xi (\phi) R$, the renormalized non-minimal coupling $\xi_R (\Phi)$ becomes
\begin{equation}
\xi_R(\Phi) \equiv \xi(\Phi)+\frac{1}{2}\xi'' (\Phi)\langle \chi^2(x) \rangle_{\text{Min}} + \frac{V''(\Phi)}{96 \pi^2} \ln \frac{\Lambda}{\WC{\mu}}
\end{equation}
and there appears a finite correction $\frac{1}{2}A(x)\xi'' (\Phi)R$ and a
logarithmically divergent term $\frac{1}{96\pi^2}\xi'' (\Phi) R^2\ln \Lambda$.
Thus, a new non-minimal coupling term of the form $\zeta (\phi)R^2$ arises, and its inclusion
in the tree-level Lagrangian yields more new terms which are higher-order in $R$.
In this paper, we assume that these non-minimally coupling terms are negligibly small and the
only non-trivial term is the (renormalized) potential. In what follows,  
we use the original symbol $V(\Phi)$ for the renormalized potential as well 
for notational simplicity. 
The third term on the right-hand side of Eq.~(\ref{1-loop-1st-order}),
which is finite and explicitly dependent on the spacetime point $x$,
cannot be absorbed into any terms at the tree level. Thus, this term is physically relevant and
contributes to the EOM of $\Phi$.

\subsection{Renormalization of the Second Order Term}
\label{Ren-2nd}
We next consider the memory term in Eq.~(\ref{effective-action-result}).
To decompose it into a divergent part and finite part, we rewrite the memory term as
\begin{equation}
\begin{aligned}
\label{memory-term}
\Gamma^{\text{(mem)}}\left[ \Phi, \Phi_{\Delta}\right] =& -\int d^4x \; a^3(t) V''(\Phi(x))V'''(\Phi(x))\Phi_{\Delta}(x)I(x)\\
& - \int d^4x \; a^3(t)V'''(\Phi(x))\Phi_{\Delta}(x)\int^t d^4x' \; a^3(t')\left[V''(\Phi(x'))-V''(\Phi(x))\right]\text{Im}\left(\brac{\chi(x)\chi(x')}^2\right),
\end{aligned}
\end{equation}
where
\begin{equation}
\label{definition-I(x)}
I(x) \equiv \int^t d^4x' \; a^3(t')\text{Im}\left(\brac{\chi(x)\chi(x')}^2\right).
\end{equation}

In order to study the potential UV divergences (i.e., $x' \to x$) in the non-local term in Eq.~(\ref{memory-term}), 
the Wightman function $W(x,x') \equiv \brac{\chi(x)\chi(x')}$ must be studied in the UV regime. 
In the limit of small invariant distance, the propagator can be expressed using the Hadamard form of the Wightman function given by \cite{decanini2008hadamard}
\begin{equation}
\label{hadamard-form}
W(x,x') = \frac{2\Delta^{\frac{1}{2}} (x,x')}{\sigma - i\epsilon} + V(x,x')\ln(\mu(\sigma-i\epsilon)) + W_{\rm reg}(x,x'),
\end{equation}
where $\sigma \equiv s^2$ is the invariant distance \footnote{
Note here that the usual definition of $\sigma$ is $\sigma \equiv \frac{1}{2}s^2$ for invariant distance $s^2$, but we adopt a different definition for computational purposes.
}
and $\Delta (x,x') \equiv -{(-g(x))}^{-\frac{1}{2}}
\det (-\sigma_{;\mu \nu' }(x,x')){(-g(x'))}^{-\frac{1}{2}}$.
The functions $V(x,x')$ and $W_{\rm reg}$ are real and finite in the limit $\sigma \rightarrow 0$.
Thus, the singular features of $W(x,x')$ are totally encoded in the first
two terms.
Expanding $\Delta$ and $V$ in the small distance limit yields
\begin{equation}
\label{small-distance-formulae}
\Delta (x,x')=1+{\cal O}(\sigma),~~~V(x,x')=v_0+{\cal O}(\sigma),
\end{equation}
where $v_0 =\frac{1}{2}m^2-\frac{1}{12}R$.
Then, with an explicit form of the singular part, $W(x,x')$ becomes
\begin{equation}
W(x,x')=\frac{1}{\sigma-i \epsilon}+v_0 \ln (\mu (\sigma-i\epsilon))+\cdots,
\end{equation}
where $\cdots$ represents a part that is regular for $\sigma \to 0$.

As we show in Appendix \ref{appendix-1}, the only UV divergent part of $I(x)$ is a logarithmic divergence from the term
\begin{equation}
     4\int^t d^4x' \; a^3(t') \times \text{Im}\left( \frac{1}{(\sigma - i\epsilon)^2}\right).
\end{equation}
This divergence arises in the case of the Minkowski space as well  
and thus it has the same divergence structure in all FLRW spacetimes.
Knowing that the only diverging part of $I(x)$ is the part that resembles the Minkowski case, the function $I(x)$ is rewritten as

\begin{equation}
\begin{aligned}
I(x) =& I_{\text{Min},\Lambda} + \left(I(x) - I_{\text{Min},\Lambda}\right) \\
= &\int^t d^4x' \; \text{Im}\left(\brac{\chi(x)\chi(x')}_{\text{Min}}^2\right)+\int^t d^4x' \; a^3(t')\left[\text{Im}\left(\brac{\chi(x)\chi(x')}^2\right)-a^{-3}(t')\text{Im}\left(\brac{\chi(x)\chi(x')}_{\text{Min}}^2\right)\right].
\end{aligned}
\end{equation}
$I_{\text{Min}}(x)$ is logarithmically divergent and independent of $x$. 
Thus, this component contributes to the renormalization of the potential at the current order.
Based on the argument above, 
\begin{equation}
B(x)\equiv I(x) - I_{\text{Min}}
\end{equation}
is finite. It varies with time on a general FLRW background and adds physical effects to
the EOM of $\Phi$. 

\subsection{Renormalized equation of motion}
Based on the \WC{work} derived in \ref{Ren-1st} and \ref{Ren-2nd}, 
\WC{by adding the counterterms
\begin{equation}
    \mathcal{L}_{\text{CT}}
    =
    \frac{1}{2}V''(\phi)(M_\Lambda+N_\Lambda)
    +
    \frac{1}{2}\left(V''(\phi)\right)^2 I_{\text{Min},\Lambda},
\end{equation}
where
\begin{equation}
    M_\Lambda =
    \frac{\Lambda^2}{8\pi^2}
    -\frac{m^2}{8\pi^2}\ln\left(\frac{\Lambda}{\mu}\right),
    \qquad
    N_\Lambda =
    \frac{R}{48\pi^2}\ln\left(\frac{\Lambda}{\mu}\right),
\end{equation}
the UV divergences appearing in the one-loop effective action within
our approximation are cancelled. The limit $\Lambda\to\infty$ is
understood to be taken after the bare parameters have been redefined
so that the renormalized quantities remain finite. As stated in
Sec.~\ref{Ren-1st}, the $\mathcal{O}(R^2)$ terms generated by the
renormalization of the non-minimal coupling parameter $\xi$ are
neglected. With this renormalization procedure, the renormalized
equation of motion becomes}
\begin{equation}
\begin{aligned}
\label{EOM-after-renormalization}
(\Box +m^2)\Phi(x) + V'(x) + \frac{1}{2}A(x)&V'''(\Phi(x)) + B(x) V''(\Phi(x))V'''(\Phi(x)) \\ +& V'''(\Phi(x))\int^t d^4x' \; a^3(t')\left[V''(\Phi(x')) -V''(\Phi(x))\right]\text{Im}\left(\brac{\chi(x)\chi(x')}^2\right)  = \xi(x).
\end{aligned}
\end{equation}
Because the factor $V''(\Phi(x'))-V''(\Phi(x))$ vanishes at $x=x'$,
the last term on the left-hand side does not generate any UV divergence.
In this way, the memory term after renormalization is split into a finite contribution to the local term and a non-local, finite memory term. 
As a result, every term in the EOM is finite and well defined.
Eq.~(\ref{EOM-after-renormalization}) is one of our main results.
This equation holds under the test-field approximation, at the 1-loop level, and 
to second order in the self-interaction.
Its validity does not rely on any assumption about the evolution of the scale factor.

Boyanovsky et al. \cite{Boyanovsky_2006} has already included the first-order correction $\frac{1}{2}A(x)V'''(\Phi(x))$ in their analysis. However, the correction to the effective potential proportional to $B(x)$, as well as the memory term and noise term, are unique to the present research. 

Owing to the presence of the memory term, the EOM (\ref{EOM-after-renormalization}) is difficult to treat analytically. 
Consequently, it is desirable to reformulate it into a more manageable approximate equation.
To this end, we note that, due to the presence of the scale factor, 
the integrand of the memory term is suppressed for $t'<t$ exponentially for nearly de Sitter space,
meaning that the memory term is dominated by the contribution from the domain near the point $x$.  
We assume that the field $\Phi$ varies slowly so that the change of the field value in that 
domain is well approximated by the first-order Taylor expansion from $x$.
This assumption, combined with the fact that the integrand is now well-behaved near coincidence, 
implies that the term $V''(\Phi(x'))$ can be approximated as (neglecting spatial derivatives of the field)
\begin{equation}
V''(\Phi(x')) \approx V''(\Phi(x)) + V'''(\Phi(x))\dot{\Phi}(x)(t'-t).
\end{equation}
Thus, the memory term can be approximated as a local term that contributes to the drift of the field as
\begin{equation}
C(x) \left(V'''(\Phi(x))\right)^2\dot{\Phi}(x) ,
\end{equation}
where $C(x)$ is defined by
\begin{equation}
C(x)\equiv \int^t d^4x' \; (t'-t)a^3(t')\times\text{Im}\left(\brac{\chi(x)\chi(x')}^2\right).
\end{equation}
With this approximation, Eq.~(\ref{EOM-after-renormalization}) can be written as
\begin{equation}
\label{EOM-local-approximation}
    (\Box +m^2)\Phi(x) + V'(x)+\frac{1}{2}A(x) V'''(\Phi(x))+B(x) V''(\Phi(x))V'''(\Phi(x)) + C(x)\left(V'''(\Phi(x))\right)^2 \dot{\Phi}(x) = \xi(x).
\end{equation}
For an FLRW spacetime, the functions $A(x), B(x)$, and $C(x)$ generally depend on time. 
If the metric is exactly de Sitter, however, they become time-independent, 
which can be seen most easily by substituting the massive mode function for $\chi(x)$ and changing variables to $\tau = t-t'$. 
This implies that in an inflationary universe, where the metric is approximately de Sitter, 
the time dependence of these functions is suppressed by the slow-roll parameters.

\begin{figure}[ht]
    \centering
    \includegraphics[width=0.9\linewidth]{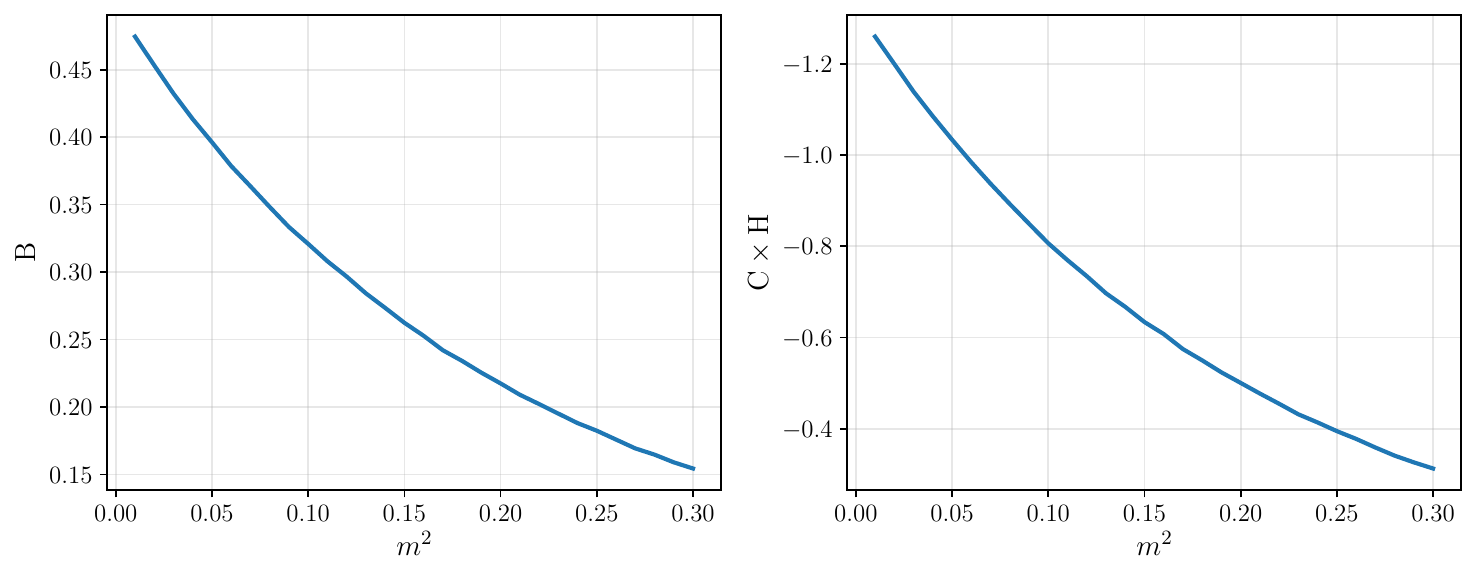}
    \caption{
    Coefficients $B$ and $C$ as functions of the mass parameter $m^2$ in units of $H^2$. 
    }
    \label{fig:BC_mass}
\end{figure}

To quantitatively evaluate $A(x), B(x)$, and $C(x)$, we assume the exact de Sitter case in the following computations.
$A$ has already been computed in \cite{Boyanovsky_2006}. 
It is shown there that $A$ scales as $\frac{H^4}{m^2}$.
Using the mode functions of the massive scalar field in de Sitter space, 
we numerically find that
\begin{equation}
\label{expression-of-A}
A \approx 0.038 \times \frac{H^4}{m^2},
\end{equation}
which is consistent with the result in \cite{Boyanovsky_2006} \WC{for $\Delta \equiv \frac{3}{2}-\nu=\frac{m^2}{3H^2} + \mathcal{O}(\frac{m^4}{H^4})$.}
$B$ is a dimensionless quantity. Since $m$ and $H$ are the only parameters present in the model,
$B$ must be a function of the ratio $m/H$.
$C$ has a dimension of time and thus $HC$ must be a function of the ratio $m/H$.
Based on this argument, values of $B$ and $C$ were numerically calculated using the massive mode function of a real scalar field in the exact de Sitter space for various values 
of $m/H$. The result is given in Fig.~\ref{fig:BC_mass}.
As the result shows, $C$ is negative, implying that the drift term acts to reduce the friction of the field. Although this result may seem counter-intuitive, it is important to keep in mind that the derivation of the equation of motion thus far has worked in a perturbative regime, where the self-interactions of the field are assumed to be small. Thus, regions where $C\left(V'''(\Phi(x))\right)^2>3H \WC{V'(\Phi(x))}$ are automatically excluded in our analysis. Thus, the friction of the field remains positive and is simply decreased slightly by the correction from $C$.

\section{Discussion}

\subsection{Connection to Stochastic Inflation}
Our formalism fundamentally differs from that of stochastic inflation, 
which models the dynamics of IR (super-horizon) modes using a Langevin equation 
driven by random noise sourced by UV quantum fluctuations. 
In the present study, we have integrated out fluctuations across all scales, 
including those exceeding the Hubble horizon.
Consequently, the variable $\Phi(x)$ in Eq. (\ref{EOM-local-approximation}) represents the quantum expectation value of the test scalar field $\phi(x)$ computed using the effective action.
Because of the interaction term $V$, the quantum fluctuations are coupled to the variable $\Phi$,
and the resulting EOM is the result of the effects from such fluctuations.\\
On the other hand, the $\Phi(x)$ that is calculated in stochastic inflation is a classical scalar field smoothed over Hubble-length scales. 
This field is given stochastic contributions arising from the continuous
transition of modes from the UV regime $(k>\epsilon aH)$ to the IR regime $(k>\epsilon aH)$
due to exponential expansion and
are present even for a free field (i.e. no $V$).
Here, $\epsilon <1$ is a small parameter dividing the two regimes \cite{Starobinsky_1994}. \\
Despite these conceptual differences, the EOM derived in this paper can be interpreted 
as a Langevin equation with radiative corrections, 
provided that contributions from IR modes ($k<\epsilon aH$) in the loop integrals 
are manually truncated and the standard stochastic noise from the UV-to-IR transition is incorporated.
Although this approach is not derived from first principles and represents 
an ad-hoc modification of our formal results, 
we expect it to serve as a reliable approximation to a full first-principles calculation.\\
Among the quantities $A, B$ and $C$ that arise from the radiative corrections, $A$ is 
particularly sensitive to IR modes because
it diverges as $m \to 0$ (see Eq.~(\ref{expression-of-A})).
Ignoring the IR contributions $(k<\epsilon aH)$ to $A$, the following replacement can be made:

\begin{equation}
    A = \int_{k=0}^{k=\infty} d^3k\;|\chi_k|^2 - \text{(UV. divergences)} \rightarrow A_{\text{sub}} \equiv \int_{k=\epsilon aH}^{k=\infty} d^3k\;|\chi_k|^2 - \text{(UV. divergences)}
\end{equation}

This results in

\begin{equation}
    A_{\text{sub}} \approx -\left(\frac{H}{2\pi}\right)^2\ln(\epsilon)
\end{equation}

Physically, this replaces variance of the fluctuations $\brac{\chi^2}$ in the local 1-loop correction with the variance of fluctuations with wavenumbers $k >\epsilon aH$.

Furthermore, our formalism, having already integrated out the fluctuations on all scales, only contains the dissipative noise term $\xi(x)$ present in Eq. (\ref{EOM-local-approximation}). Thus, we artificially include a stochastic white noise term with amplitude $\frac{H}{2\pi}$ to the EOM. 

With the above modifications, the EOM (\ref{EOM-local-approximation}) is reducible to the Langevin equation

\begin{equation}
\label{Langevin-eq1}
\dot{\Phi}
=
-\frac{V'_{\mathrm{eff}}(\Phi)}{3H}
\left(
1 + \frac{C}{3H}\left(V_{\text{eff}}'''(\Phi)\right)^2
\right)^{-1}
+ \xi(x) + \xi_0(x),
\end{equation}
where the noise now includes the one $\xi_0$ present in standard stochastic inflation
and we have introduced the {\it effective} potential $V_{\rm eff}$ by
\begin{equation}
V_{\text{eff}} =\frac{1}{2}m^2\Phi^2 + V + \frac{A_{\text{sub}}}{2}V''+\frac{B}{2}\left(V''\right)^2.
\end{equation}
As the correlation function of $\xi$ is given by Eq.~(\ref{noise-correlation}),
the amplitude of $\xi$ is proportional to $V'''$ and the noise is colored.
On the other hand, the correlation function of $\xi_0$ is given by \cite{Starobinsky_1994}
\begin{equation}
\brac{\xi_0(\xv, t)\xi_0(\xv,t')} = \frac{H^3}{4\pi^2}\delta(t-t').
\end{equation}
As this equation shows, $\xi_0$ is white noise whose amplitude is not suppressed by $V'''$. 
Thus, when the self-interaction is weak, $\xi_0$ provides the dominant contribution, 
and we neglect the noise $\xi$ in the following. 
A further technical reason for ignoring $\xi$ is that the standard Fokker-Planck equation
does not apply in the presence of colored noise and incorporating such noise. An explicit treatment of colored noise using influence-functional methods has been explored in related contexts, see \cite{Lee_2011}.
With this approximation, the Fokker-Planck equation corresponding to the Langevin 
equation (\ref{Langevin-eq1}) can be obtained in the standard manner as
\begin{equation}
\label{Fokker-Planck-eq}
    \frac{\partial P(\Phi)}{\partial t} = \frac{1}{3H} \frac{\partial}{\partial \Phi}\left[\frac{V'_{\text{eff}}}{1+\frac{C}{3H}(V_{\text{eff}}''')^2} \times P(\Phi)\right]+\frac{H^3}{8\pi^2}\frac{\partial^2P(\Phi)}{\partial \Phi^2}.
\end{equation}

As an application of Eq.~(\ref{Fokker-Planck-eq}), we consider a simple example where the renormalized potential is given by $V(\phi) = \frac{\lambda}{4}\phi^4$, where $\lambda$ is a dimensionless parameter. Note that the field is massive, and the mass term is excluded from the potential.
After a sufficiently long time, the probability distribution $P(\Phi)$ settles down to the 
equilibrium configuration $P_{\rm eq}$ which is time independent.
This configuration is obtained as a static solution of Eq.~(\ref{Fokker-Planck-eq})
and is given by 
 \begin{equation}
 \label{equildis}
 P_{\rm eq}\left(\Phi\right) = N \exp \left(\frac{8\pi^2}{3H^4}\left[ \ln \left( 1 + \gamma \Phi^2 \right)^{\mu} + \frac{\beta \Phi^2}{2\gamma}\right]\right),
 \end{equation}
where $N$ is the normalization constant and 
\begin{equation}
    \mu \equiv \frac{\alpha\gamma - \beta}{2\gamma^2} \; , \; \alpha \equiv -m^2 -6\lambda A_{\text{sub}} \; , \; \beta \equiv -\lambda - 18\lambda^2B \; , \; \gamma \equiv 12\lambda^2(1+18B\lambda)^2 \frac{C}{H}.
\end{equation}

\begin{figure}[t]
    \centering
    \includegraphics[width=0.7\linewidth]{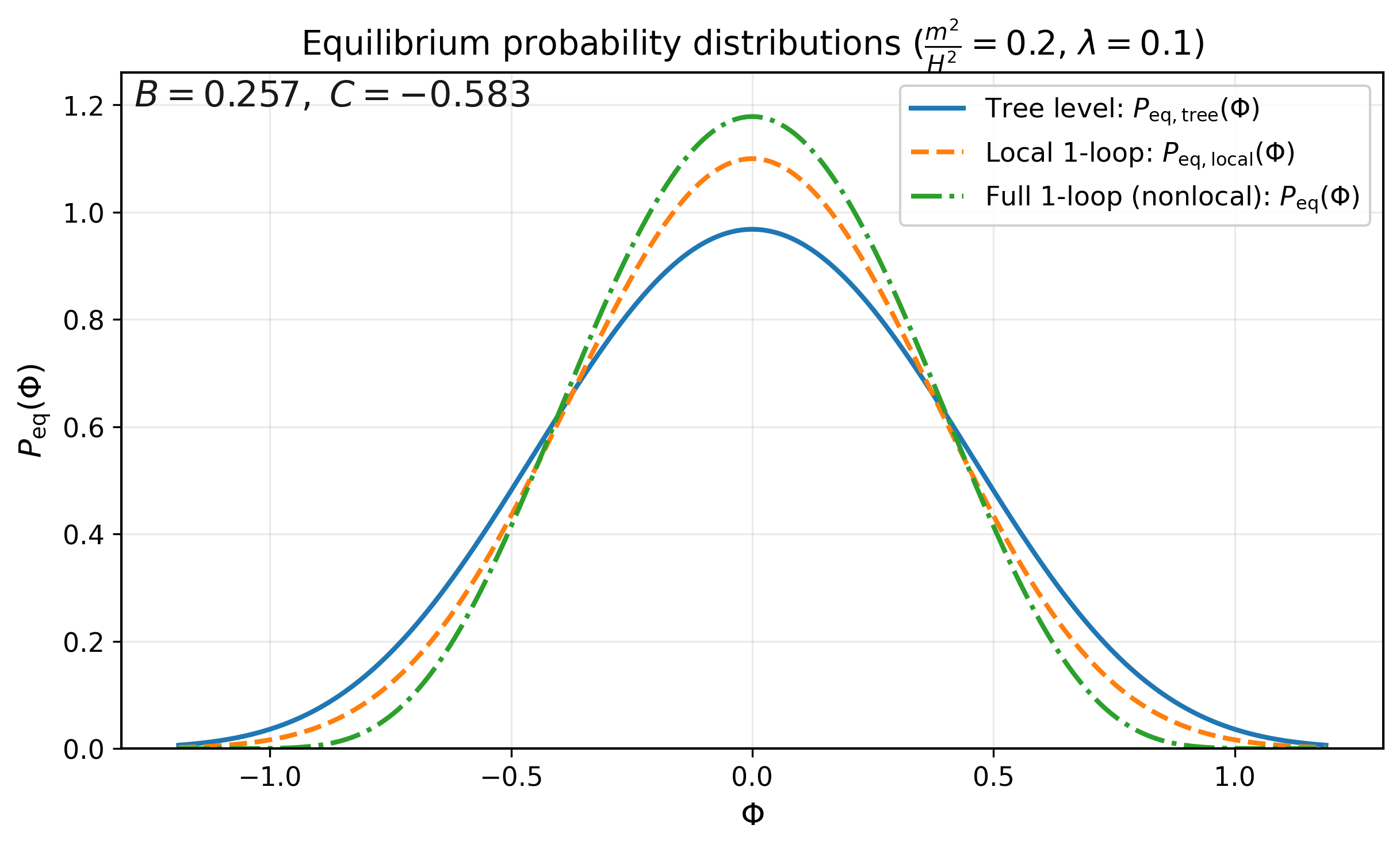}
    \caption{A comparison of the equilibrium probability distribution for three cases: tree level, the first order correction, and the full non-local corrections. $\Phi$ is given in units of $H$.}
    \label{fig:equildists}
\end{figure}

Fig.~\ref{fig:equildists} shows the equilibrium probability distribution of the full non-local case, as well as the local first order case ($B+C=0$) and the tree level result. 

In the limit of small $\Phi$, the natural logarithm can be expanded, and the equilibrium distribution can be expressed as:
\begin{equation}
    P_{\rm eq}\left(\Phi\right) = N\times P_{\rm eq, tree} (\Phi)\times \exp\left[-\frac{8\pi^2}{3H^4}\left[6\lambda A\Phi^2 + \left(-18\lambda^2 \frac{BCA_{\text{sub}}}{H}+3m^2\lambda\frac{C}{H}-\frac{9}{2}\lambda^2\right)\Phi^4 + \mathcal{O}(\lambda^3)\right]\right],
\end{equation}
where $P_{\rm eq,tree}$ is the equilibrium configuration at the tree level.
Thus, we see that near the center of the distribution the one-loop corrections serve to change the effective values of the mass and the coupling constant.

\begin{figure}[t]
    \centering
    \includegraphics[width=0.9\linewidth]{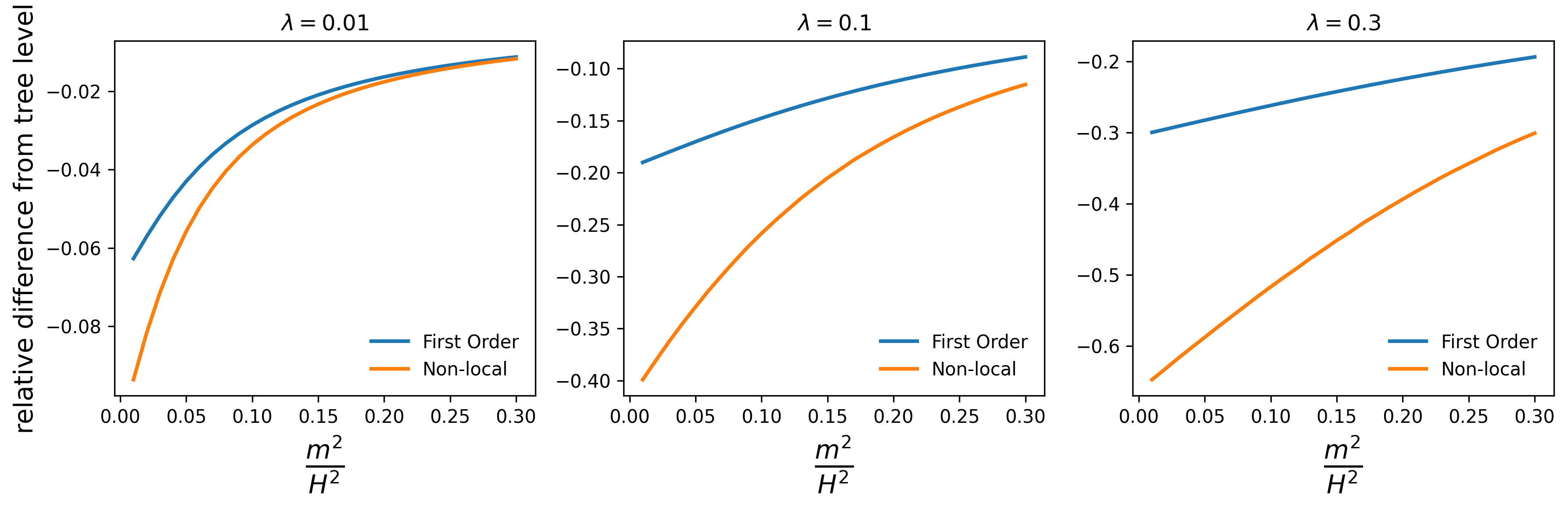}
    \caption{The relative difference of the variance from the tree level result as a function of $\frac{m^2}{H^2}$ for three different values of the coupling constant $\lambda$.}
    \label{fig:phi2}
\end{figure}

Fig.~\ref{fig:phi2} shows the value of $\brac{\Phi^2}$ as a function of mass for two cases: the local one-loop corrected case  and the full, non-local one-loop corrected case. Since the only action of $A$ is to contribute to the effective mass of the field, the equilibrium probability distribution for the local case can easily be read off as

\begin{equation}
    P_{\rm eq,local}\left(\Phi\right) = N \times \exp \left[- \frac{8\pi^2}{3H^4}\left(\frac{1}{2}\left(m^2 + 6\lambda A_{\text{sub}}\right)\Phi^2 + \frac{\lambda}{4}\Phi^4\right)\right].
\end{equation}

As mentioned earlier, our work restricted to potentials $V$ which allow perturbative expansion. This in turn restricts the allowed values of $\Phi$ in the case of massive $\phi^4$ theory.
Looking at Eq.~(\ref{equildis}), values of the field that cause $1+\gamma\Phi^2 < 0$ are forbidden as they would assign complex values to the probability distribution. Thus, this provides a natural upper limit for the value of the field. Specifically,
$$1+\gamma \Phi_{\text{max}}^2 = 0 \rightarrow  \Phi_{\text{max}} \approx \frac{H}{\lambda}\alpha,$$

where $\mathcal{O}(\lambda^3)$ corrections have been neglected and $\alpha \equiv \sqrt{-\frac{1}{12CH}}$ is an $\mathcal{O}(1)$ dimensionless constant.

Our results confirm that the size of $\brac{\Phi^2}$ is largely determined by the effective mass of the field. The behavior of the graphs in Fig.~\ref{fig:phi2}  can be explained by the fact that the effective mass is roughly 
\begin{equation}
    m^2_{\text{eff}} \approx m^2 + 6\lambda A_{\text{sub}}.
\end{equation}
 Thus, stronger coupling will lead to stronger restoring forces in both the mass and interaction term, and thus decrease the variance of the field. This matches with the numerical results shown in Fig.~\ref{fig:phi2}.

 $B$ serves to increase the effective coupling strength of the field, which, along with $A$, increases the restoring force and sharpens the equilibrium distribution. The effective coupling constant is

 \begin{equation}
     \lambda_{\text{eff}} = \lambda + \frac{9}{2}\lambda^2B.
 \end{equation}

\begin{figure}[H]
  \centering
  \includegraphics[width=0.9\textwidth]{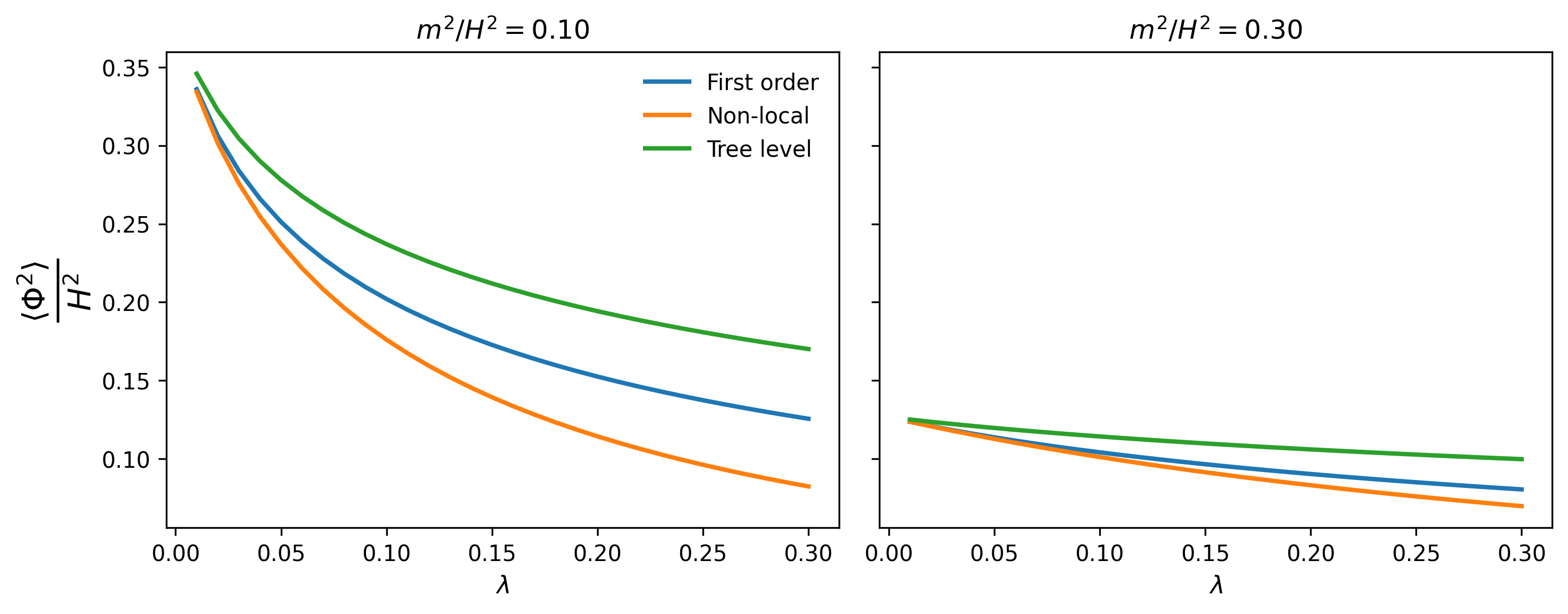}
  \caption{
    Expectation value $\frac{\brac{\Phi^2}}{H^2}$ as a function of $\lambda$.
    Both panels show first-order and non-local corrections.
  }
  \label{fig:phi2_mass_comparison}
\end{figure}

It is of interest to see how the non-local contribution grows as a function of $\lambda$ for fixed values of $m^2$. Fig.~\ref{fig:phi2_mass_comparison} shows the growth of the separation between the tree result, the one-loop result with only local corrections, and the result with non-local corrections for masses $m^2 = 0.1H^2 \; , \; 0.3H^2$.

The variance decreases as a function of $\lambda$, which is in alignment with the fact that the effective mass and effective coupling constant increase with an increase in $\lambda$. Also, the contribution from the second order terms grows with stronger coupling which explains the increase in the separation between the first order and non-local result.
\subsection{Corrections to the Noise}

In this paper we aimed to derive a Fokker-Planck equation in order to compare the equilibrium probability distribution and expectation value of the variance to standard tree-level stochastic inflation. In order to do this we neglected the dissipative noise with correlation
$$\brac{\xi(x)\xi(y)} = \frac{1}{2} a^3(t_x) a^3(t_y) V'''(\Phi(x))V'''(\Phi(y))\text{Re}\biggr( \brac{\chi(x)\chi(y)}^2\biggr)$$ 
since the Fokker-Planck equation only applies to a Langevin equation with white noise. One of the other reasons for neglecting the noise was the suppression of the amplitude of the noise by $V'''$ compared to the standard amplitude of the UV modes.
Despite this, using the analytic expression of the variance, general statements about the contributions to the motion of the field from the dissipative noise can be made. Considering a stochastic process of the field at spacetime point $x$, the variance is given by:
$$\sigma^2 = \brac{\xi^2(x)} = \frac{1}{2}a^3(t_x)(V'''(\Phi(x)))^2\brac{\chi^2(x)}^2$$
Assuming the noise sourced from dissipative effects and from the UV modes are uncorrelated, this implies the one-loop noise term contributes positively to the strength of the drift. Since the non-local effects on $\brac{\Phi^2}$ consistently decrease the variance of the equilibrium probability distribution as shown in Fig. \ref{fig:equildists}, it is likely that the noise has some mitigating effects that partially cancel the effects of the memory term.
In order to fully study the effects of dissipative noise on the present system, a more elaborate treatment that properly incorporates the colored noise is required.

\section{Conclusion}

In this paper, we have investigated the quantum corrections to the dynamics 
of a test scalar field in a general FLRW background by computing the one-loop effective action. 
By employing the Schwinger-Keldysh (in-in) formalism and expanding the action 
perturbatively up to the second order in the interaction, 
we have derived the equation of motion for the expectation value of the field.

We found that the resulting equation of motion contains not only the standard 
local radiative corrections to the potential, 
which are derived in previous studies \cite{andersen2021quantum, Kamenshchik_2022, Markkanen_2018, Boyanovsky_2006}, 
but also novel nonlocal terms such as a memory term and a stochastic noise term. 
These nonlocal features emerge naturally at the second order of the interaction. 
We have successfully identified all UV divergent structures within these terms 
and demonstrated that the equation of motion can be consistently reformulated into a renormalized, finite form. 
The memory term has been shown explicitly to respect causality as it only incorporates information from the past light cone.

\addednew{We expanded to second-order in the spacetime-dependent mass-like term $V''(\Phi(x))$ (mass insertions) in our calculation of the effective action, but in principle it is desirable to evaluate Eq~(\ref{1-loop-action}) analytically (mass resummation). We leave this task for future work.}

To address the analytical complexity introduced by the nonlocal memory term, 
we applied a local expansion of the field under the assumption that the field varies slowly over time. Under this approximation, 
we found that the memory term manifests as a correction to the drift term in the equation of motion. 
Our numerical analysis revealed that this correction provides a negative contribution 
to the drift coefficient, effectively reducing the friction experienced by the field.

Finally, we applied this formalism to a massive $\phi^4$ theory as a concrete example. 
Our results indicate that the inclusion of these one-loop corrections leads to a suppression 
of the field variance in the IR regime compared to the tree-level result.

Before closing this section, it should be emphasized that the present study 
focuses on a test scalar field, 
and thus the results cannot be directly applied to the inflaton field. 
Since the inflaton dominates the energy density of the universe, 
a complete derivation of the inflaton’s equation of motion requires the inclusion 
of not only the inflaton's fluctuations but also the metric perturbations. 
We leave incorporating these gravitational degrees of freedom for future work.

\section*{Acknowledgments}
We would like to thank Tomotaka Kuroda for his helpful comments.
This work was supported by JSPS KAKENHI Grant Number JP23K03411 (TS).

\appendix
\section{Convergence Properties of $I(x)$}
\label{appendix-1}
The purpose of this appendix is to clarify the UV divergence of the quantity
$I(x)$ defined by Eq.~(\ref{definition-I(x)}).
Substituting in the Hadamard form (\ref{hadamard-form}), we have
\begin{equation}
    I(x) =\int^t d^4x'\; a^3(t')\;\text{Im} \left[\left(\frac{2\Delta}{\sigma-i\epsilon} + V\ln \left(\mu(\sigma-i\epsilon\right) + W_{\text{reg}}\right)^2\right].
\end{equation}
Any terms containing $W_{\rm reg}$ give only finite contributions to $I(x)$, and we therefore
ignore $W_{\rm reg}$ in what follows.
Furthermore, it is safe to keep only the leading order term in $\Delta$ and $V$ in the expansion formulae (\ref{small-distance-formulae}).
We also expand $\sigma$ in the small distance as
\begin{equation}
\sigma=-{(t-t')}^2+a^2(t)r^2+\cdots,
\end{equation}
where $r=|{\bm x}-{\bm x'}|$ is the comoving distance and $\cdots $ is the higher-order part which is ignored in the following calculations.
Thus, the UV part of $I(x)$ can be written as
\begin{equation}
\label{definition-IUV}
I_{\rm UV}(x)=\int^t d^4x'\; a^3(t')\;\text{Im} \left[\left(\frac{2}{-{(t-t')}^2+a^2(t)r^2-i\epsilon} + v_0\ln \left(\mu(-{(t-t')}^2+a^2(t)r^2-i\epsilon\right)\right)^2\right].
\end{equation}
Expanding the square, three terms appear as potential sources of UV divergence. 
As we will show below, one of these terms contains a logarithmic divergence in the limit of coincident points, and all others converge in the UV regime.

\subsection{Evaluation of the first term}
The term in $I_{\rm UV}(x)$ we evaluate first is given by
\begin{equation}
     I_{\rm UV,1}(x)= 4\int^t d^4x'\; a^3(t')\;\text{Im}  \left[\left(\frac{1}{-{(t-t')}^2+a^2(t)r^2-i\epsilon}\right)^2\right].
\end{equation}
This corresponds to the square of the first term in the outer parentheses in Eq.~(\ref{definition-IUV}).
Switching to polar coordinates, changing the variables as
$\tau=t-t', R=a(t)r$, and using the following identity
\begin{equation}
    \text{Im} \left[\left(\frac{1}{x-i\epsilon}\right)^2\right] = -\pi \delta'(x)
\end{equation}
gives
\begin{equation}
   I_{\rm UV,1}(xs)= -16\pi^2 \int_0 d\tau \frac{a^3(t-\tau)}{a^3(t)}\int dR\; R^2 \delta'\left(R^2 -\tau^2\right).
\end{equation}
Making the additional change of variable by $s=R^2$,
the above expression becomes 
\begin{equation}
    I_{\rm UV,1}(x)=4\pi^2 \int_0 d\tau \; \frac{a^3(t-\tau)}{a^3(t)} \; \frac{1}{\tau}.
\end{equation}
This is logarithmically divergent near $\tau\equiv t-t'=0$, and since the delta function was activated for $R^2=\tau^2$, this divergence corresponds to coincident points.
As it is obvious in the final expression of $I_{\rm UV,1}$, there is no
effect of the scale factor at $\tau=0$ and the divergence
takes the same form as that in the Minkowski space. 

\subsection{Evaluation of the second term}
The term in $I_{\rm UV}(x)$ we consider next is given by
\begin{equation}
I_{\rm UV,2}(x) =  4\pi v_0^2\text{Im} \left[ \int_0 d\tau \frac{a^3(t-\tau)}{a^3(t)} \int dR~R^2 
\ln^2 (\mu(-\tau^2+R^2 - i\epsilon)) \right],
\end{equation}
where the same change of variables as that used for $I_{\rm UV,1}$ has been already performed.
This corresponds to the square of the second term in the outer parentheses in Eq.~(\ref{definition-IUV}).
Rewriting the natural logarithm as
\begin{equation}
\label{natural-log}
\ln(\mu(\sigma - i\epsilon)) = \ln|\mu(\sigma - i\epsilon)| + i \text{arg}(\mu\sigma - i\epsilon),
\end{equation}
we have
\begin{equation}
I_{\rm UV,2}(x) =  8\pi v_0^2 \int_0 d\tau \frac{a^3(t-\tau)}{a^3(t)} \int dR~R^2 
\ln |\mu(-\tau^2+R^2 - i\epsilon)| \times
\text{arg}(\mu(-\tau^2+R^2 - i\epsilon))
\end{equation}
Because the integrand only diverges logarithmically in the UV,
its integration is convergent.

\subsection{Evaluation of the third term}
The term in $I_{\rm UV}(x)$ we consider last is given by

\begin{equation}
    I_{\rm UV,3}(x)= 4v_0\int_0 d\tau \frac{a^3 (t-\tau)}{a^3(t)}\int dR~R^2\; \text{Im} \left[\frac{\ln\left(\mu(-\tau^2+R^2-i\epsilon)\right)}{-\tau^2+R^2-i\epsilon}\right].
\end{equation}
This corresponds to the cross term of the first and the second terms in the outer parentheses in Eq.~(\ref{definition-IUV}).

Using the distributional identity 
\begin{equation}
\frac{\ln\!\left[(\mu(\Delta x^2-\epsilon))\right]}{\Delta x^2-i\epsilon}
=
\frac{1}{8}\partial^2
\left\{
\ln^2\!\left[(\mu(\Delta x^2-i\epsilon))\right]
-2\ln\!\left[(\mu(\Delta x^2-i\epsilon))\right]
\right\},
\end{equation}
the imaginary kernel of $I_{\text{UV,3}}(x)$ can be rewritten as
\begin{equation}
    \text{Im} \frac{\ln(\mu(\Delta x^2 -i\epsilon))}{(\Delta x^2 -i\epsilon)} = \frac{1}{8}\partial^2W[R,\tau], \; \; \; \; \; W[R,\tau] \equiv-2\pi(\ln(|\mu\Delta x^2|)-1)\Theta(\tau-R).
\end{equation}
Then, the integral can be written as
\WCC{
\begin{equation}
I_{\mathrm{UV},3}(x)
=
-\pi v_0 \partial_x^2
\int_0^{\infty}\frac{d\tau}{a^3(\tau)}
\int_0^\tau dR \;R^2
\left[
\ln\left(\mu|\tau^2-R^2|\right)-1
\right].
\end{equation}
Using
\begin{equation}
\int_0^\tau dR \;R^2
\left[
\ln\left(\mu|\tau^2-R^2|\right)-1
\right]
=
\frac{\tau^3}{3}
\left[
\ln(4\mu\tau^2)-\frac{11}{3}
\right],
\end{equation}
one obtains, in exact de Sitter spacetime,
\begin{equation}
I_{\mathrm{UV},3}(x)
=
\pi v_0
\lim_{L\to\infty}
\left(
\frac{d}{dt}
\left[
e^{-3Ht}
\frac{t^3}{3}
\left(
\ln(4\mu t^2)-\frac{11}{3}
\right)
\right]_{t=L}
\right).
\end{equation}
Since the exponential factor dominates the polynomial and logarithmic factors,
\begin{equation}
I_{\mathrm{UV},3}(x)
=
\pi v_0
\lim_{L\to\infty}
\left(
\frac{d}{dt}
\left[
e^{-3Ht}
\frac{t^3}{3}
\left(
\ln(4\mu t^2)-\frac{11}{3}
\right)
\right]_{t=L}
\right)
=0 .
\end{equation}
Thus this term does not introduce an additional UV divergence and no additional counterterm is required.
}

\bibliography{mybib}{}
\bibliographystyle{unsrt}

\end{document}